# P-Selectivity, Immunity, and the Power of One Bit[*]

Lane A. Hemaspaandra
Department of Computer Science
University of Rochester
Rochester, NY 14627
USA

Leen Torenvliet
ILLC
University of Amsterdam
1018 TV 24 Amsterdam
The Netherlands

April 25, 2005; revised December 7, 2005

## Abstract

We prove that P-sel, the class of all P-selective sets, is EXP-immune, but is not EXP/1-immune. That is, we prove that some infinite P-selective set has no infinite EXP-time subset, but we also prove that every infinite P-selective set has some infinite subset in EXP/1. Informally put, the immunity of P-sel is so fragile that it is pierced by a single bit of information.

The above claims follow from broader results that we obtain about the immunity of the P-selective sets. In particular, we prove that for every recursive function $f$, P-sel is DTIME($f$)-immune. Yet we also prove that P-sel is not $\Pi_2^p/1$-immune.

## 1 Introduction

This paper studies whether the class of P-selective sets is so complex as to be immune to various uniform and nonuniform classes. A set $B$ is *P-selective* exactly if there is some polynomial-time computable function $h$ such that, for each $x$ and $y$, it holds that (a) $h(x,y) \in \{x,y\}$ and (b) if either $x$ or $y$ belongs to $B$ then $h(x,y) \in B$ ([Sel79]; among the other early key papers that started the study of P-selectivity are [Sel81,Sel82a,Sel82b, Ko83]). There are many reasons for studying the P-selective sets. Among those reasons are that the P-selective sets can give insight into the relative power of polynomial-time reductions, the P-selective sets are related to heuristic search in which at each stage one merely wants to know "between these two alternatives, which is more likely to succeed," the study of selectivity can provide insights into seemingly unrelated areas (the nondeterministic analog of P-selectivity was used to show that if NP has unique solutions then the polynomial hierarchy collapses [HNOS96]), the P-selective sets are the most intensely studied of a broad range of classes of "partial information" classes [NT03], and the P-selective sets are particularly interesting in that they seem to have conflicting results on their complexity—in some ways they are complex and in some ways they are easy. For a far more detailed presentation of the motivations for studying the P-selective sets, see [HT03, Preface].

[*]Supported in part by grant NSF-CCF-0426761. Work done in part while L. Hemaspaandra was visiting the University of Amsterdam. Conference version [HT06] (Thirty-second International Conference on Current Trends in Theory and Practice of Computer Science).

Let us turn to the last-mentioned theme. Are the P-selective sets complex or are they easy? In some senses they are known to be simple, e.g., if SAT is P-selective then P = NP, the P-selective sets are in the extended low hierarchy, and the P-selective sets all have small circuits (belong to P/poly). In some senses, they are known to be hard, e.g., every tally set (no matter how difficult) polynomial-time Turing reduces to some P-selective set.

The focus of the present paper is on whether the P-selective sets are so hard as to be immune to standard uniform and nonuniform complexity classes. A class $\mathcal{C}$ is said to be *immune* to a class $\mathcal{D}$ exactly if there is some infinite set in $\mathcal{C}$ that has no infinite subset that belongs to $\mathcal{D}$. Informally put, no $\mathcal{D}$ set can recognize an infinite number of the elements of $\mathcal{C}$'s "difficult" set without incorrectly claiming that nonelements of that set belong to that set—in some sense, $\mathcal{D}$ cannot "approximate well from the inside" the hard set. Immunity is a so-called "strong separation," and clearly $\mathcal{C}$ being $\mathcal{D}$-immune always implies that $\mathcal{C}-\mathcal{D} \neq \emptyset$ (see Figure 1a). (This paper provides an example for which the converse fails. Though such failure is not common, at least one example is widely known: Not all r.e. sets are recursive, but each infinite r.e. set has an infinite recursive subset. Another—admittedly relativized—example is that with probability one relative to a random oracle the high levels of the boolean hierarchy separate from the boolean hierarchy's second level [Cai87], yet regardless of the oracle, each infinite set in the boolean hierarchy contains an infinite subset from the second level of the boolean hierarchy [CGH$^+$88].)

It is very often the case in complexity theory that when two classes can be (absolutely or with some oracle) separated, then they can be (absolutely or with some other oracle) separated with immunity. Nonetheless, we will show that such an extension is impossible in some cases regarding the P-selective sets. In particular, it is known that P-sel $\not\subseteq$ EXP/$n$ (although it also is known that P-sel $\subseteq$ NP/$n + 1$) [HT96]. Nonetheless, this result cannot possibly be extended to immunity, as we note that every infinite P-selective set has an infinite $\Pi_2^p/1$ (and thus certainly EXP/1) subset.

So, as mentioned above, P-sel is not immune to sufficiently powerful nonuniform classes.

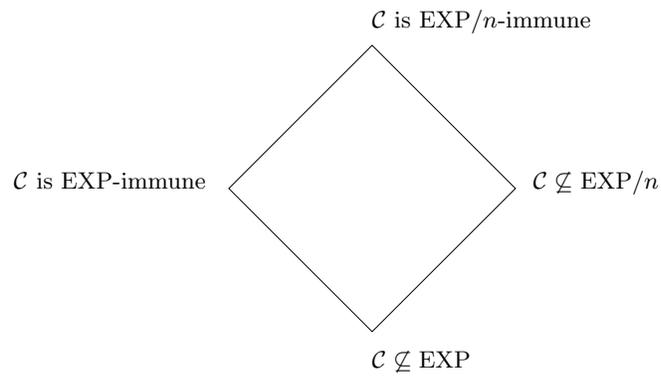

Figure 1a: Lattice of separation strengths comparing a class $\mathcal{C}$ with EXP and EXP/$n$

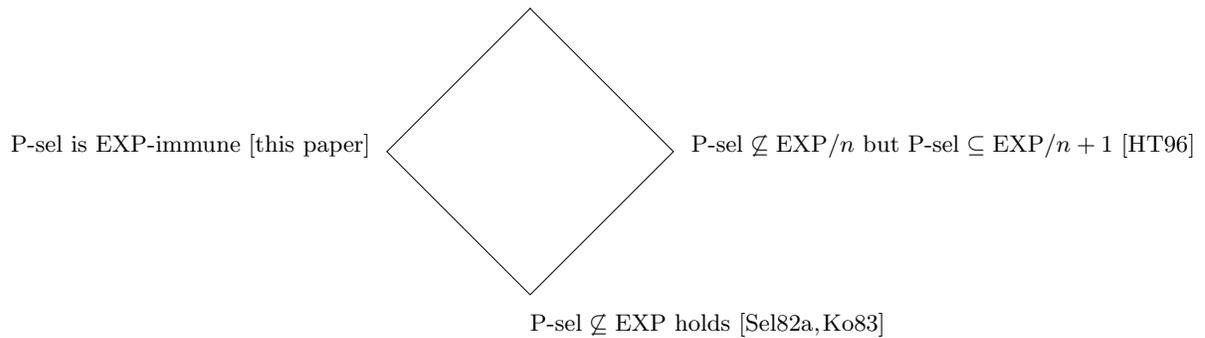

Figure 1b: P-sel versus EXP, in the context of separation-vs-immunity and $n$ or $n+1$ advice bits

Nonetheless, we show that P-sel *is* immune to any nice time-bounded class. In particular, we show that for any recursive function $f$, P-sel is DTIME($f$)-immune.

In fact, we will show a bit more. Building on the result of [HT96] that for any recursive function $f$, P-sel is not contained in DTIME($f$)$/n$ (a result that this paper shows cannot possibly be extended to immunity), Hemaspaandra, Hempel, and Nickelsen [HHN04] showed that for any recursive function $f$, A-P-sel is not contained in DTIME($f$)$/n$ (also a result that this paper shows cannot possibly be extended to immunity), where A-P-sel is the class of all sets that are P-selective via some selector function that itself is associative (it is known that all such sets are contained in P$/n+1$ [HHN04]). We will show that for any recursive function $f$, P$/1 \cap$ A-P-sel is DTIME($f$)-immune.

Thus, from our results it follows that in many cases the immunity of the P-selective set hinges on a single bit (per length). For example, P-sel is EXP-immune but P-sel is not EXP$/1$-immune. We know of no other natural examples of standard classes whose immunity relative to a class is pierced by the very first extra bit of information.

## 2 Definitions

Recall that a set $B$ is *P-selective* exactly if there is some polynomial-time computable function $h$ such that, for each $x$ and $y$, it holds that (a) $h(x,y) \in \{x,y\}$ and (b) if either $x$ or $y$ belongs to $B$ then $h(x,y) \in B$ [Sel79]. We say that such a function $h$ is a P-selector function for $B$. P-sel is defined as $\{A \mid A$ is P-selective$\}$. If $B$ is P-selective via some P-selector function that is associative, we way that $B$ is *associatively P-selective* (equivalently, $B \in$ A-P-sel) [HHN04].

As mentioned in the introduction, a class $\mathcal{C}$ is said to be *immune* to a class $\mathcal{D}$ exactly if there is some infinite set in $\mathcal{C}$ that has no infinite subset that belongs to $\mathcal{D}$ (see [Rog67], and introduced into complexity theory at least as early as [Ber76], see also [SB84]).

As is standard, $\Pi_2^p$ denotes coNP$^{\text{NP}}$ and EXP denotes $\bigcup_{k>0}$ DTIME($\mathcal{O}(2^{n^k})$).

Karp–Lipton advice classes—which capture the power of a given complexity class when augmented by a certain number of bits of free, nonuniform "advice" at each length—are defined in the standard way ([KL80]; for uniformity, we take the wording of the following definition directly from [HO02]). In particular, phrased very informally, a class $A/g$ captures the power of the class $A$ when it is helped, on each input string $x$, by being given a $g(|x|)$-bit "advice" string whose value depends only on $|x|$.

**Definition 2.1**    *1. For any set $A$ and any function $g$, $A/g$ denotes the class of all sets*

$L$ such that for some function $r$ satisfying $(\forall n)\,[|r(n)| = g(n)]$ it holds that

$$L = \{x \mid \langle x, r(|x|)\rangle \in A\}.$$

2. For any class $\mathcal{C}$ and any function $g$, $\mathcal{C}/g$ denotes

$$\{L \mid (\exists C \in \mathcal{C})\,[L \in C/g]\}.$$

## 3 Results

As mentioned in the introduction, it is known that for any recursive function $f$, P-sel (and even A-P-sel) is not contained in $\mathrm{DTIME}(f)/n$ [HT96, HHN04]. The following two results show that this nonuniform result cannot possibly be extended to immunity, but that its uniform analog does extend to the case of immunity.

**Theorem 3.1** *For each recursive function $f$, P-sel (and indeed even P/1 $\cap$ A-P-sel) is $\mathrm{DTIME}(f)$-immune.*

**Theorem 3.2** *P-sel is not $\Pi_2^p/1$-immune.*

**Corollary 3.3** *If $\mathrm{P} = \mathrm{NP}$, then P-sel is not $\mathrm{P}/1$-immune.*

**Corollary 3.4** *P-sel is not $\mathrm{EXP}/1$-immune.*

We will prove Theorem 3.2 first, as it is the easier to prove.

**Proof of Theorem 3.2** Consider an arbitrary infinite P-selective set $B$. Let $h$ be a P-selector function for $B$. Without loss of generality, we may assume that $h$ is commutative—otherwise replace it with $h(\min(x,y), \max(x,y))$, which remains a P-selector for $B$ and is clearly commutative. At each length $n$, consider the tournament induced by this selector function on the strings in $\Sigma^n$. By the induced tournament, we mean the graph on the strings in $\Sigma^n$ such that between each pair of distinct strings $x$ and $y$ there is an edge in exactly one direction, namely, if $f(x,y) = x$ then the edge points from $x$ to $y$. We say a node $w$ of a tournament is a *king* if each node in the tournament can be reached from $w$ via paths of length at most two. By an old result of Landau, each nonempty tournament has a king [Lan53]. Hemaspaandra, Ogihara, Zaki, and Zimand [HOZZ] proved that testing whether a given string is a king of a tournament can be done in $\Pi_2^p$.[1] Note that if any string

---

[1] Indeed, though we do not need these facts here, king-testing in tournaments in fact is $\Pi_2^p$-*complete* [HHW05], and kings have also played a key role in proving the first-order definability of—and thus

at length $n$ belongs to $B$, then certainly each king of length $n$ belongs to $B$. Thus, infinite set $B$ has an infinite $\Pi_2^p/1$ subset. In particular, our $\Pi_2^p/1$ set works as follows: The one bit regarding length $n$ says whether $B \cap \Sigma^n \neq \emptyset$, and our $\Pi_2^p$ set, on input $\langle z, b \rangle$, $z \in \Sigma^*$, $b \in \{0, 1\}$, accepts exactly if advice bit $b$ is one and the string $z$ is a king in the tournament induced by $f$ on the strings in $\Sigma^{|z|}$. ❑

We now turn to our main result, namely, Theorem 3.1: For each recursive function $f$, P-sel (and indeed even A-P-sel) is DTIME($f$)-immune.

As mentioned earlier, for each recursive function $f$, P-sel (and even A-P-sel) has been separated previously, though not with immunity, from DTIME($f$) [HT96,HHN04]. Each of those proofs works by direct diagonalization. Unfortunately, due to the fact that machines can "play possum" (can for long periods of time accept no strings), direct diagonalization does not seem to work for the more demanding case of separation with immunity. To handle this, we differ from those previous papers, and create immunity, by employing an injury-free "waiting"/priority-type construction. Our construction is a rather unusual one in that, though some requirements may remain active and unsatisfied forever, we ensure that at each stage we simultaneously satisfy all active requirements that can be satisfied via acting just at the current, relevant length. In short, we give a "waiting"/priority-type argument that is so egalitarian as to not need explicit priorities. While doing so, we integrate the strengths of each of the earlier proofs mentioned above, namely, we retain the gap-like approach of [HT96] and part of the scheme of [HHN04] for assuring associativity, and for purposes of clarity, comparison, and connection, where possible we retain as far as we can their notations and arguments. However, note that our proof creates a simpler framing structure: By putting all strings at a length in or out as a block, we avoid not just explicit priorities, but also we avoid the issues of "left cut"ing (or in [HHN04], "right cut"ing) that are central to the earlier work, since (unlike those earlier proofs that truly needed them) here they would just be needlessly entangling the proof and argument.

**Proof of Theorem 3.1** Let $f$ be a recursive function. Then there will always exist a strictly increasing recursive function $f'$ such that: (a) for each $n$, $f'(n) > \max(f(n), 2)$, and (b) for some Turing machine $M$ that computes $f'$ it holds that for each $n$ the machine computes $f'(n)$ using at most $f'(n)^{\mathcal{O}(1)}$ steps. Fix such a function $f'$.

---

upper-bounding the complexity of—the reachability problem for certain nonsuccinctly specified restricted graph types [NT02]. Also, we mention that [HOZZ] itself proves some immunity claims, most particularly that P-sel is not bi-immune to the class of weakly-P-rankable sets, and that P-sel is not immune to the class of weakly-FP$^{\Sigma_2^p}$-rankable sets. However, due to the nature of the notion of weak-rankability (which we do not define here), those nonimmunity results seem to be of no help regarding obtaining the current result.

Let $\ell_1 = 2$. For each $i \geq 1$, let $\ell_{i+1} = 2^{2^{2^{2^{f'(\ell_i)}}}}$. Let $\mathcal{L} = \{\ell_1, \ell_2, \ldots\}$. Our set $B$ will have the following properties: (a) $B \in \text{DTIME}(2^{2^{f'(n)}})$, (b) $x \in B \implies |x| \in \mathcal{L}$, and (c) $(|x| = |y| \wedge x \in B) \implies y \in B$.

We must and will also ensure that $B$ is an infinite set.

Note that any set satisfying (c) is in P/1, since at each length it will contain either all strings or no strings, and so one bit of advice easily suffices to accept the set. So, in our particular case, $B \in \text{P}/1$. (In fact, even the P interpreter is overkill. The complexity is even lower, at least if one's pairing function from the definition of advice classes is such that it doesn't really require the full power of P to decode its subparts.)

Any set $B$ satisfying (a), (b), and (c) is associatively P-selective. This follows by an argument given (though not in the explicit context of the $f'$-based gaps we use above) in [HHN04], and for completeness, we argue that $B$ is in A-P-sel. Let $\widehat{\mathcal{L}}$ denote $\{x \mid |x| \in \mathcal{L}\}$. Our selector function for $B$ will be

$$h(x,y) = \begin{cases} \max(x,y) & \text{if } ||\{x,y\} \cap \widehat{\mathcal{L}}|| = 0 \\ \{x,y\} \cap \widehat{\mathcal{L}} & \text{if } ||\{x,y\} \cap \widehat{\mathcal{L}}|| = 1 \\ \min(x,y) & \text{if } ||\{x,y\} \cap \widehat{\mathcal{L}}|| = 2 \wedge |x| = |y| \\ \min(x,y) & \text{if } ||\{x,y\} \cap \widehat{\mathcal{L}}|| = 2 \wedge |x| \neq |y| \wedge \min(x,y) \in B \\ \max(x,y) & \text{if } ||\{x,y\} \cap \widehat{\mathcal{L}}|| = 2 \wedge |x| \neq |y| \wedge \min(x,y) \notin B. \end{cases}$$

In the second line here, by $\{x,y\} \cap \widehat{\mathcal{L}}$ we, in a slight abuse of type rules, mean the unique element in the named 1-element set, rather than the set itself. It is clear that $h$ is in P since, if $||\{x,y\} \cap \widehat{\mathcal{L}}|| = 2 \wedge |x| \neq |y|$, then—since $B \in \text{DTIME}(2^{2^{f'(n)}})$ and the lengths in $\widehat{\mathcal{L}}$ are quadruple-exponentially spaced—we given $x$ and $y$ can brute-force test "$\min(x,y) \in B$?" in time polynomial in $|\max(x,y)|$. It is also clear that $h$ is a P-selector function for $B$. To see that $h$ is associative, first notice that $h$, which is clearly commutative, has the following properties. When both of $h$'s inputs belong to $B$, $h$ outputs the (lexicographically) smaller input. When exactly one input belongs to $B$, $h$ outputs that one. And when neither input belongs to $B$, $h$ outputs the lexicographically smallest input whose length equals the largest value in $\mathcal{L}$ that is the length of one of the inputs if any input belongs to $\widehat{\mathcal{L}}$, and otherwise $h$ outputs the larger input. In light of these observations, it is clear that when for any $a$, $b$, and $c$ the function $h$ is applied twice in any of the 12 possible ways (six possible orderings of $a/b/c$ and two possible groupings for each, though due to commutativity some are instantly seen as the same), the result is the same, namely, if at least one of $a/b/c$ belongs to $B$ then the result is the lexicographically smallest string in $\{a,b,c\} \cap B$. If none of $a/b/c$ belong to $B$, then the result is the lexicographically smallest string of length $\max(\{|a|,|b|,|c|\} \cap \mathcal{L})$ if $\{|a|, |b|, |c|\} \cap \mathcal{L} \neq \emptyset$, and otherwise is $\max(a, b, c)$. So $h$ is associative.

Let $M_1$, $M_2$, $M_3$, ... be a simple enumeration of Turing machines such that this enumeration has the properties that (a) $\bigcup_k L(M_k) \supseteq \text{DTIME}(f')$, and (b) for each $k$ and each $y$, the running time of machine $M_k$ on input $y$ is at most $2^{kf'(|y|)}$ steps. And let our enumeration be such that each language in $\text{DTIME}(f')$ is accepted by infinitely many machines from this enumeration (this keeps us from having to worry about problems if a few small values of $k$ might seem to cause tension with respect to our $B \in \text{DTIME}(2^{2^{f'(n)}})$ constraint).

We now turn to the stage construction that will define $B$. Our construction proceeds in stages. Let requirement $R_i$ be defined as "$L(M_i) \not\subseteq B$." Once we start trying to satisfy a requirement, it will be said to be active (until we mark it, if ever, as satisfied).

At stage $k$, we will define which strings of length $\ell_k$ are in $B$. At stage $k$, we will have as active all requirements $R_1$, ..., $R_{k/2}$, other than those that we have marked as being satisfied.

At the start of stage $k$, we will rebuild the history of this construction, and in doing so, will determine which strings are in $B$ at each interesting (i.e., member of $\widehat{\mathcal{L}}$) length of $B$ that is strictly less than $\ell_k$, and—rather crucially—will also determine which requirements were, during that part of the construction, marked as being satisfied. (Since this involves objects whose length is upper-bounded by a quadruple-logarithm of the length of our input, it is not hard to see that this will not interfere with the goal of ensuring that $B \in \text{DTIME}(2^{2^{f'(n)}})$.) After doing so, we seek to extend $B$ at the current length, $\ell_k$, in such a way as to satisfy at least one active requirement. So, for each $M_i$ corresponding to an active requirement, see if there is any string of the current length that the machine accepts. Let $I$ denote the set of all $M_i$'s corresponding to active requirements for which there is such a string. If $I \neq \emptyset$, we will leave $B$ at the current length completely empty, and for each $i \in I$ we will mark $R_i$ as satisfied ($M_i$ accepts a string that is not in $B$, so $M_i$ certainly does not accept a subset of $B$).

On the other hand, if $I = \emptyset$ then we will put into $B$ all strings of of length $\ell_k$. Note that since, through stage $k$, we have activated no more than $k/2$ requirements, this case ($I = \emptyset$) happens at least half the stages, and so our set $B$ will indeed be an infinite set.

Note that each requirement eventually becomes active. There are two cases. If in some stage after the point where it becomes active the requirement is one of those that can be satisfied, then it be marked as satisfied then. Otherwise, it (say, $R_i$) indeed will never be marked as satisfied, but that means—since there were only finitely many stages that occurred before the point where $R_i$ became active—that $L(M_i)$ contains only a finite number of strings at lengths in the set $\mathcal{L}$, and so certainly is not an *infinite* subset of $B$. Thus, for each $i$, we know that it will be the case that $L(M_i)$ cannot be an infinite subset of $B$. ❑

We start to wind up by giving an example of how these theorems can be applied to a concrete class. Theorem 3.1 on its surface applies to a class defined by a single time-function, $f$. Note that for natural time classes, the class is defined by an infinite collection of functions that in general have no single function that majorizes them (since each may have bigger and bigger constants). Nonetheless, to inherit our result it suffices to majorize each almost everywhere, and so our result is inherited by all standard time-bounded and space-bounded classes. For example, consider EXP. From the fact that P-sel is immune to DTIME($2^{2^n}$) it is easy to see that P-sel is immune to EXP. This holds because if a set $B$ has an infinite subset in EXP, it clearly (regardless of what huge constants may apply to the EXP algorithm) has an infinite subset in DTIME($2^{2^n}$), e.g., by having the time $2^{2^n}$ machine accept no strings at all except when $n$ has become so large that $2^{2^n}$ is so much bigger than the exponential bound of the EXP machine that our machine can easily simulate the EXP machine. So, we have that P-sel is clearly EXP-immune (and, for that matter, EEEEEXP/immune too). Combining this observation with the known result that P-sel is contained in EXP/$n+1$ (even NP/$n+1$) but is not contained in EXP/$n$ [HT96], Figure 1b now as an example presents (in light of Corollary 3.4) what regular and strong separations hold for EXP with respect to $n$ and $n+1$ bits of advice.

On the other hand, we leave as an open issue whether Theorem 3.1 can be extended to show that P-sel is RECURSIVE-immune. Note that our result shows something weaker, namely, that for each recursive function $f$, P-sel is immune to DTIME($f$); though trivially

$$\text{RECURSIVE} = \bigcup_{\{f \,|\, f \text{ is a recursive function}\}} \text{DTIME}(f),$$

our result nonetheless does not imply that P-sel is RECURSIVE-immune, and the proof technique does not seem to generalize to yield that.

Finally, we mention that our result is not really specific to P-selectivity, but that the natural analogs exist for other selectivity types (based on the corresponding complexity for generating and evaluating king-ness in their induced tournaments), e.g., EXP-sel is immune to DTIME($f$) for each recursive function $f$, and yet EXP-sel is not EXP/1-immune (as one can, still within EXP-time, brute-force king-finding in tournaments based even on EXP-selectors).

**Acknowledgments** We thank Piotr Faliszewski, Xiao Zhang, and the referees of the Thirty-second International Conference on Current Trends in Theory and Practice of Computer Science for helpful comments.

[Sel82b] A. Selman. Reductions on NP and P-selective sets. *Theoretical Computer Science*, 19(3):287–304, 1982.